\documentclass[11pt,aps,prb]{revtex4}
\usepackage[latin1]{inputenc} 
\usepackage [dvips]{graphicx}  
\newcommand{\ee}{\end{equation}} 
\newcommand{\be}{\begin{equation}} 
 
\newcommand{\ba}{\begin{array}}                   
\newcommand{\ea}{\end{array}} 
 
\begin{document} 
\title{From discrete to continuous dynamics and back: How large is 1?}
\author{Hubert Krivine$^{(a)}$, Annick Lesne$^{(b)}$,  Jacques 
Treiner$^{(a)}$\footnote{Email addresses for correspondence:
krivine@ipno.in2p3, lesne@lptl.jussieu.fr, treiner@ccr.jussieu.fr}}
\affiliation{(a) LPTMS, B\^atiment 100,
 Universit\'e Paris-Sud, F-91405 Orsay}
\affiliation{(b) LPTL, Case 121, Universit\'e Pierre et Marie Curie,
4 place Jussieu, F-75252 Paris.}

\begin{abstract}
Discrete autonomous dynamical systems in dimension 1
 can exhibit chaotic behavior, whereas the
corresponding continuous evolution equations rule it out,
and cannot even possess a nontrivial periodic solution.
 Therefore the passage from
discrete to continuous equations (and conversely)
is all  but harmless. We address this issue and evidence
some caveats on the paradigmatic
 Verhulst  logistic equation, investigating in particular
the status and influence of the actual size of the unit time
step in  discrete modelings, rooted in
well-known numerical analysis.
\end{abstract}
 
\maketitle


\section{The discrete-time logistic evolution}

The logistic map $f_a(x)=ax(1-x)$
 giving the celebrated recursion relation
on the interval $[0,1]$ \be 
x_{n+1}=ax_n(1-x_n)=f_a(x_n)
\hskip 10mm x_0\in[0,1] \hskip 5mm a\in ]1,4]
\label{e:1}\ee is one of  the simplest 
example of discrete autonomous evolution leading to chaos. This 
nonlinear
equation was introduced by Verhulst
(a Belgian mathematician)
in 1838  [\cite{Verhulst}]
to take into account that $a$, the Malthus coefficient 
characterizing the growth of the population
$$X_{n+1}=aX_n,$$ has to decrease when $X_n$ increases, due to
 resources limitation. The simplest way was to replace the
constant rate $a$ by a linear dependence in $X_n$,
matching the rate $a$ at vanishing population, namely
$a(1-X_n/M)$; the parameter $M$ is then interpreted as
being the maximum acceptable population,
currently known as the ``carrying
capacity'' of the environment.
 Equation (\ref{e:1}) is recovered through
the change of variable
 $x_n=X_n/M$.
A very rich variety of dynamic behaviors
is generated by Eq. (\ref{e:1}), whose
  temporal structure 
is governed  by the values of the control parameter $a$.
Since the seminal 
reference [\cite{MAY76}], several studies of the asymptotic
dynamics of  (\ref{e:1}) have been published, among which some
very pedagogical ones are [\cite{PJS}] and [\cite{KJ}].
 Let us only recall the most significant properties.

For $a$ given, such that 
$1<a<a_1=3$, the fixed point  $x^*_a=1-1/a$ is stable, globally 
attractive, therefore $x_n\to x^*_a$  as $n\to\infty$,  irrespectively 
of the initial condition $x_0$ provided it belongs to its
 basin of attraction $]0,1[$. 
In $a_1=3$, a cycle of  period 2 appears
through a pitchfork bifurcation. Also called
period-doubling bifurcation since it is associated with
 the destabilization of a fixed point $x^*_a$ into a 2-cycle
(or the destabilization of a $2^n$-cycle into a $2^{n+1}$-cycle
when it involves $f^{2^n}_a$ instead of $f_a$), this  
  generic bifurcation is 
characterized by the relation
$f'_{a_1}(x^*_{a_1})\equiv \partial_xf(a_1,x^*_{a_1}) =-1$
 and the generic condition
$\partial_{ax}
f(a_1,x^*_{a_1})\neq 0$ (denoting here the $a$-dependence
on the same footing for the sake of clarity) [\cite{iooss}].
The 2-cycle emerging in $a_1$ remains stable and globally attractive in $]0,1[$
for any $a<a_2=1+\sqrt{6}$.  More 
generally, there exists an increasing
 sequence  $(a_k)_k$ of bifurcation values 
such that  for $a_k<a<a_{k+1}$, the asymptotic regime  is a cycle of 
period $2^k$, which destabilizes
in $a_{k+1}$ through a pitchfork
bifurcation of $f_a^{2^k}$. 
 This sequence converges to  $a_{\infty}\approx 3.5699$ 
according to the  scaling law $a_{\infty}-a_k\sim \delta^{-k}$  with 
a universal rate $\delta\approx 4.6692$ \ [\cite{Feig}].
The discrete evolution  (\ref{e:1})  is actually a generic 
example exhibiting this so-called
 period-doubling scenario towards chaos, i.e.  a 
normal form to which any one-parameter family experiencing such a 
scenario is conjugated [\cite{CE}].
In 
$a=a_{\infty}$,  a chaotic behavior arises, reflecting
for $a>a_{\infty}$ in a 
positive Lyapounov exponent   (sensitivity to 
initial  conditions) and mixing property
(dynamic decorrelation of phase space regions).
  Chaotic regions in the 
$a$-space then intermingle in a  highly complicated fashion (but now 
understood [\cite{CE}]) with non chaotic
 regions where  stable odd cycles rule the 
asymptotic dynamics.

The  conclusion, now acknowledged but striking at the time of
 publication of Ref. [\cite{MAY76}] and anyhow remarkable,
is that a large variety of chaotic behaviors 
can be generated by a one-dimensional discrete evolution, 
with a seemingly harmless nonlinearity
(smooth and simply quadratic). 
It showed that nonlinearities are never harmless when 
supplemented with a folding dynamics, here coming 
from the bell shape of the evolution map. 
 

\section{Continuous-time counterpart: a trivial dynamics} 
As it is impossible
to give an analytical solution of (\ref{e:1}), 
i.e.  $x_n$ as an explicit function of $n$ and $x_0$, and because we 
are interested in the asymptotic solution  $n\to\infty$
(which gives a vanishing relative duration to the unit step
$n\to n+1$)
 it is 
appealing   to deal with the corresponding continuous 
problem [\cite{Hubbard}],  which is straightforwardly
 solvable. 
To derive  a continuous 
 counterpart of  (\ref{e:1}), one subtracts $x_n$ to 
 both sides of equation (\ref{e:1}) 
and identifies  $x_{n+1}-x_n$ with the differential of a continuous 
function of time $y(t)$, which leads: 
\be 
\frac{dy}{dt}=f_a(y)-y=y[a(1-y)-1],\label{e:2}\ee whose analytical 
solution is easily obtained : \be 
y(t)=\frac{(a-1)y_0}{ay_0+[a(1-y_0)-1]e^{-(a-1)t}}.\label{e:3}\ee 
This solution is obviously regular
with respect to $t\geq 0$ for any value of $a>1$ and, not surprisingly, 
tends to $x^*_a$ when $t\to\infty$.
In contrast with this plain behavior, qualitatively insensitive
to the value of $a>1$,  any attempt to solve 
(\ref{e:2}) by discretization with a time step $h=1$ will lead to the logistic 
evolution (\ref{e:1}) with its full richness of solutions 
as $a$ is varied. On the other hand one expects that, 
for $h$ small enough, 
one should approach the true solution (\ref{e:3}). 
How is it possible ? We 
have therefore to quantify what means ``small enough''.

\section{Interpretation of discretization schemes
 associated with the logistic equation}

Let us thus recall the behavior of the discretization schemes
associated with (\ref{e:2}) \cite{borrelli}.
Our aim is evidently not
to get more knowledge about this equation,
nor to device an accurate numerical resolution,
but rather to understand
in this tractable and well-understood situation what is currently
done to solve real problems when no straightforward
solution is available.

For a given time step 
$h$, the discretization scheme  writes 
\be \label{e:4a}y(t+h)=y(t)+h\{ay(t)\left[1-y(t]\right)-y(t)\}\ee
A remarkable feature of the logistic equation is the 
possibility to rewrite this scheme as 
 \be Y(t+h)=AY(t)\left(1-Y(t)\right),\label{e:4}\ee with 
\be  Y(t)=\lambda y(t) 
\hskip 3mm{\rm where }\hskip 3mm \lambda={ah\over 1+a(h-1)}\ee 
involving the effective control parameter
\be
A(a,h)=1+h(a-1)\ee
provided $y_0\in [0,1/\lambda]$ (note that $\lambda<1$ if $h<1$).
Obviously, the same 
 phenomenology as for evolution (\ref{e:1}) will be observed. For instance, 
 the inequality $A<3$, required to obtain the convergence 
 of (\ref{e:4}) to 
 the nontrivial fixed point $Y^*_A=1-1/A$,  means 
 \be
h<h_c(a)= \frac{a_1-1}{a-1}={2\over a-1}\ee 
Extending the reasoning to the subsequent bifurcations, one would 
observe a period-doubling scenario when the
discretization  step $h$ increases, 
namely at values $(h_k)_k$ with $A(a, h_k)=a_k$, i.e. 
\be h_k={a_k-1\over a-1}\ee 
Chaos arises for $h>h_{\infty}(a)=(a_{\infty}-1)/(a-1)$. 
The bifurcation diagram  as a function of $h$, at fixed $a$, would then 
be similar 
to the standard bifurcation diagram in $a$-space, up to a rescaling of the  
attracting sets by a factor of $\lambda(a,h)$,
 and a translation and rescaling of the bifurcation values
($a_k=1+(a-1)h_k$). 
In particular, it is interesting to  
note that the sequence $(h_k)_k$ follows the same universal 
scaling law $h_{\infty}-h_k\sim \delta^{-k}$ 
or more precisely: 
\be\frac{h_{i+1}-h_i}{h_{i+2}-h_{i+1}}\longrightarrow \delta 
\hskip 3mm 
{\rm when}\hskip 3mm i\to\infty 
\hskip 3mm 
{\rm with} \hskip 3mm\delta\approx 4.4669\ee 
For illustration let us consider the case $a=3.1$ 
(Figures 1, 2 and 3). 
The critical value 
 of $h$ is $h_c=(a_1)/(a-1)=2/2.1\simeq 0.9524$.
For $h>h_c$, one sees a 2-cycle, namely 
 oscillations of the 
 solution  between the two (stable) fixed 
 points of $f_A[f_A(Y)]$. 
The onset of the chaos is for $h=h_\infty=(a_{\infty}-1)/(a-1)=
2.5699/2.1=1.22376$.


\section{Discussion: an interplay between two characteristic times}
 This simple study illustrates that the passage
 from
continuous to discrete nonlinear equation is not insignificant:
destabilization  of the continuous time evolution, 
leading to cycles and even a spurious chaotic behavior,
follows from an improper choice of the step of the discretization
[\cite{Matano}] or conversely an actual chaotic behavior can be
suppressed by replacing a discrete model by its
limiting  continuous counterpart.


\vskip 3mm
Nevertheless, the passage from equation (\ref{e:1}) to (\ref{e:4}) by a simple
scaling is exact only in the case of the quadratic family. We
shall now  investigate what remains in more general
situations.
Let $f$ be a map, generating a discrete dynamical system 
$x_{n+1}=f(x_n)$ and having  a stable  fixed point $x^*$
(i.e. $f(x^*)=x^*$ and $|f'(x^*)|<1$). 
The naive continuous counterpart writes 
$dy/dt=f(y)-y$.
Linear stability analysis
shows  that $x^*$ is still a  (at least locally) stable
fixed point of the continuous dynamics since
the linear growth rate of perturbations is negative: $f'(x^*)-1<0$.

 We might then consider the discrete scheme
$z_{n+1}=z_n+h[f(z_n)-z_n]$ for various values of 
the time step $h$. It is straightforward to show that this 
discretization scheme destabilizes for 
$h>h_c$ where
\be h_c=2/[1-f'(x^*)]\ee 
Indeed, the linear stability of $x^*$ breaks down when the modulus
$|1+h(f'(x^*)-1)|$ overwhelms 1, which occurs for
$1+h(f'(x^*)-1)=-1$.
 This relation yields the above value of $h_c$ and shows that
 the discrete scheme exhibits a period-doubling (pitchfork) bifurcation 
in $h=h_c$
 (the  additional generic condition for this bifurcation 
stated in Section 1
being also fulfilled, as can be directly checked).

\vskip 3mm
The additional feature observed when  the map 
$f_a$ depends on a control parameter $a$ and 
 exhibits a period-doubling 
in $a_1$ is that $h_c(a)$ crosses $h=1$ in $a=a_1$: 
for $a>a_1$, $f'_a(x^*_a)<-1$ and 
$x^*_a$ is instable with respect to the 
initial  discrete dynamics ($h=1$) 
 but is still a stable fixed point of the continuous dynamics,
showing the inadequacy of the limiting continuous model $dy/dt=f_a(y)-y$ to 
capture the behavior of the
discrete one $x_{n+1}=f_a(x_n)$.
 It is to note that $h_c(a)$ decreases if $a$ increases: the more stable 
is the fixed point (i.e. the larger $|f'_a(x^*_a)-1|$
with   $f'_a(x^*_a)-1<0$), the smaller is the time-step range of validity 
of the discretization scheme (in a sense, the less stable is the 
discretization scheme).

\vskip 3mm
The qualitative differences, explicitly described in the previous
sections, between the continuous-time and 
discrete-time versions of the logistic
equation (and above in a more general
framework)  are not really surprising: a general claim assesses 
that a continuous-time dynamics requires a phase space of dimension
at least 3 to develop a chaotic behavior [\cite{Hilborn}].
In dimension 1 or 2, continuous trajectories behave
as boundaries each for each other (trajectories of an autonomous continuous
dynamic system cannot cross each other), which obviously prevents from
chaos (and even nontrivial periodic solutions in dimension 1). 
But whereas it is thus straightforward to foresee
the loss of chaotic and even periodic behavior
 when turning to the limiting continuous
dynamics, is it possible to understand on physical grounds
 the existence of a critical value $h_c$ for the 
discretization time step $h$? The explanation lies in the comparison
of the intrinsic time scale(s) of the dynamics
with the chosen ``time unit'' $h$. 

The characteristic time of a continuous evolution, still denoted
$dy/dt=f(y)-y$ to avoid proliferation of new notations,
can be estimated as $\tau\sim 1/[1-f'(x^*)]$.
Indeed, a mere linearization of (\ref{e:2}) around the fixed
point $x^*$ leads to:
\be
{d\over dt} [x(t)-x^*]=[f'(x^*)-1](x-x^*)\ee
 hence
the value of $\tau$.
Destabilization of the discretization scheme
occurs when $h>h_c=2\tau$.
The stepwise updating, after each time step $h$, of the evolution law is
too rough to properly control
the discrete evolution and force it to follow closely all
the relevant variations of the continuous trajectory.
This is reminiscent of the Nyquist theorem [\cite{NYq}] for a periodic 
continuous evolution: the observation time step should 
be smaller than half the smallest period 
(or characteristic time) to properly 
sampling the continuous trajectory. 

It is to note that $\tau$ or equivalently the critical value $h_c=2\tau$
of the time step are  intrinsic features of the dynamics, in the sense that
they are invariant through conjugacy: for any diffeomorphism
$\phi$, $f(y)$ and $\phi^{-1}\circ f\circ \phi(y)$
(providing an equivalent modeling of the discrete model associated with $f$) 
 or $f(y)$ and $y+\phi^{-1}[f\circ \phi(y)-\phi(y)]$
(providing an equivalent modeling of the continuous model
 associated with $f(y)-y$) will have the same critical value $h_c$
 and the same characteristic time $\tau$.

Let us carry further the
comparison between the continuous evolution and its discretization,
in order  to understand the emergence of oscillations for $h>2\tau$.
The general continuous equation $dy/dt=f(y)-y$
 operates a fine tuning
of the evolution rate $dy/dt$ that is obviously not achieved
 by updating $f(y)-y$ at times $t_n=nh$. We have shown
here that, near a stable fixed point,
  the resulting discrepancies lead to a bifurcation in the asymptotic
dynamics, when $h$ overwhelms the characteristic time of the evolution.
To take a familiar example of such oscillations arising 
from a mismatch between two characteristic times, let us consider an
heating/cooling device, able to measure 
the difference between
the instantaneous room temperature
and a prescribed one, and to monitor the appropriate
 energy supply or extraction,
to compensate the measured difference.
If the time $h$ necessary for the device to actually deliver
the required energy is longer that the characteristic time of 
temperature variations in the environment, the device will
not balance the external temperature variations but rather,
its ill-phased response will superimpose and the room temperature
will suffer large oscillations.
More generally, any ill-tuned homeostatic device, responding with
a large lag  $h$, will produce oscillations, and the  result
of Section 3 is the mathematical translation of this ubiquitous phenomenon.

\section{Conclusion}

In conclusion, we have presented 
an example showing explicitly the link  between the validity 
of the discretization scheme 
with the dynamical (in)stability of the associated map for a unit step-size. 
Convertely, it enlights the specificity of the discrete dynamics, 
that cannot in general be understood, even qualitatively, from 
the behaviour of its continuous counterpart. 
In two or more dimensions,  an additional problem arises: : the 
recursion relation is no more  unique [\cite{FUT}].  
The caveats illustrated in this paper are all the more relevant.


\newpage
\begin{figure}[h]
\begin{center}
\includegraphics[angle=-90,scale=.4]{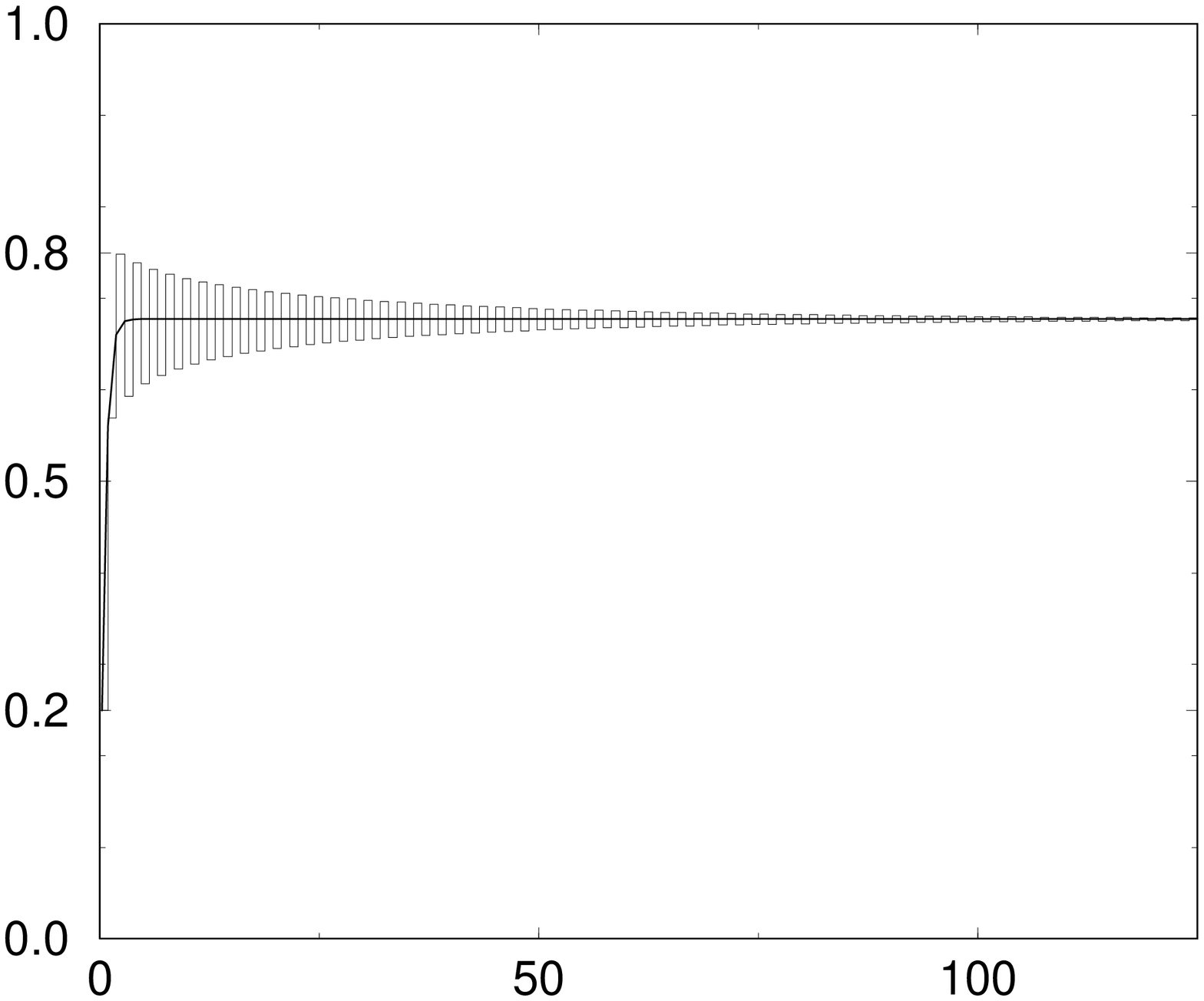}
\caption{\it Discretization
of the logistic equation (\ref{e:2}) with $a=3.1$,
 using a time step $h<h_c$
(here $h=0.94$ whereas $h_c\equiv 2/(a-1)=20/21\approx 0.95$),
 see text, Section 3.
Bold line: exact (continuous-time) solution
of (\ref{e:2}). Stair step $\frac{1}{\lambda}f_A(nh)$.
$x_a^*=1-1/a$.}
\end{center}
\end{figure}

\newpage
\begin{figure}[h]
\begin{center}
\includegraphics[angle=-90,scale=.4]{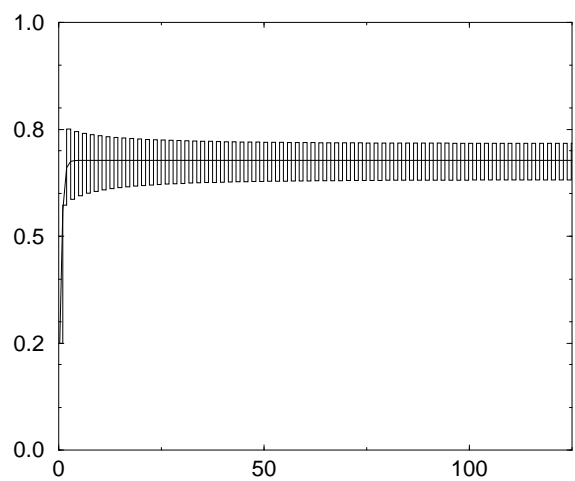}
\caption{\it Same as Fig. 1 but with $h=0.96>h_c$.}
\end{center}
\end{figure}

\newpage
\begin{figure}[h]
\begin{center}
\includegraphics[angle=-90,scale=.4]{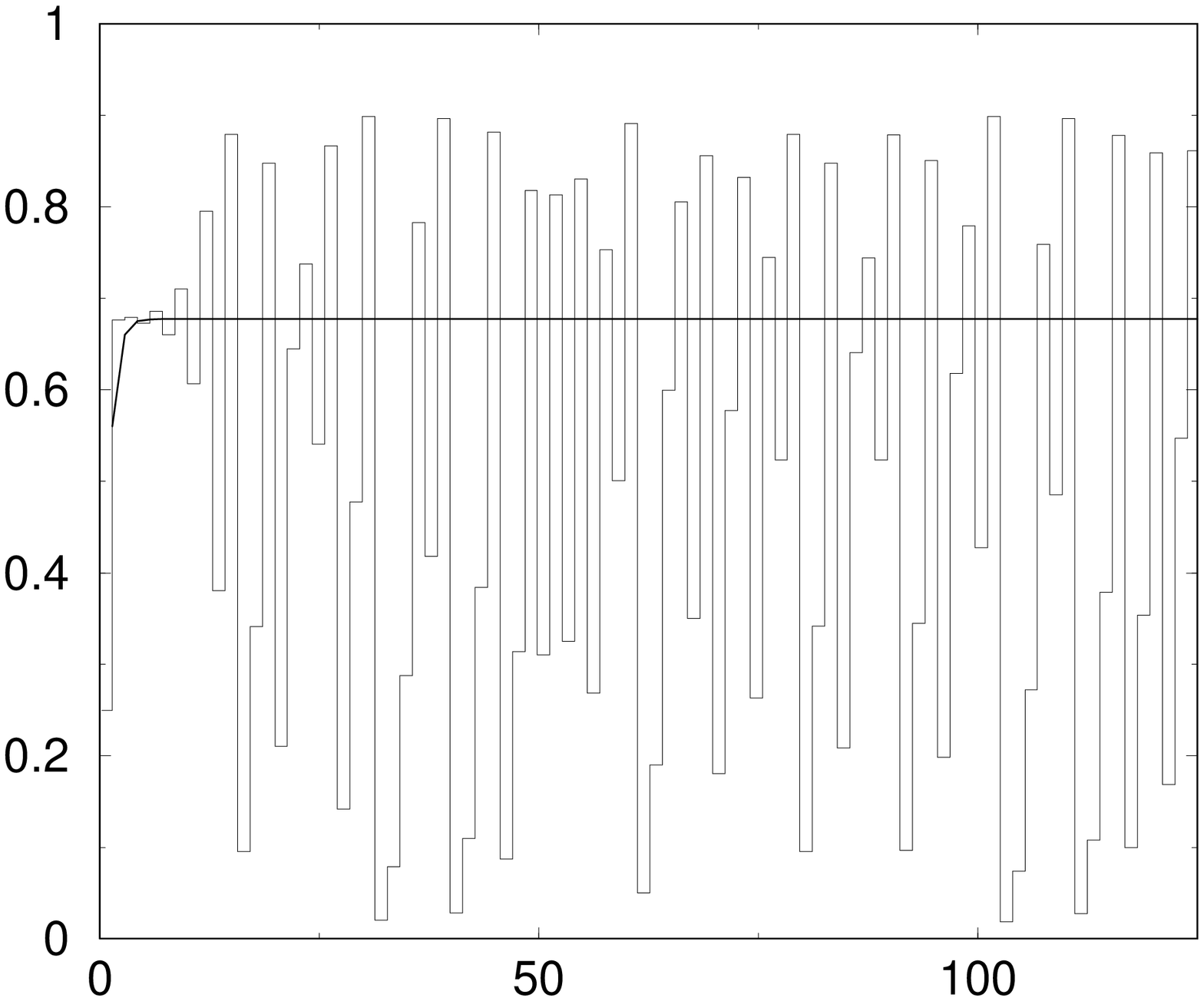}
\caption{\it  Discretization of the logistic equation 
(\ref{e:2}) with $a=3.1$, using a time step $h=3/(a-1)=1.424$
corresponding to the fully chaotic case $A=4$, see text, Section 3,
and [\cite{a4}].}
\end{center}
\end{figure}


\begin{thebibliography}{99}
\bibitem{a4}
Except for $a=4$, where 
$x_n=\sin^2(2\pi\theta_n)$ with 
$\theta_n=2\theta_{n-1}=2^n\theta_0$
if $x_0=\sin^2(2\pi\theta_0)$. This equivalence with
the angle-doubling $\theta_{n+1}=2\theta_n$  (modulo 1) allows
to prove that one gets a fully chaotic behavior  for $a=4$
(the location of $x_n$ below 1/2, coded 0,  or above 1/2, coded 1, 
generates a binary sequence that is  
statistically equivalent to the outcome of a game  of 
heads-and-tails).
\end{thebibliography}
\end{document}